\documentclass[english,12pt]{article}

\usepackage{physics} 
\usepackage[symbol]{footmisc} 
\usepackage{hyperref} 
\hypersetup
{colorlinks   = true, 
 urlcolor     = blue,
 linkcolor    = blue, 
 citecolor    = blue}

\usepackage{cancel} 

\numberwithin{equation}{section}
\usepackage{mathtools}
\usepackage{tensor}
\usepackage{array}
 \usepackage[normalem]{ulem}
\usepackage{graphicx}
\usepackage{amssymb}
\usepackage[english]{babel}
\usepackage{amsmath}
\usepackage{multirow}
\usepackage{prettyref}
\usepackage{babel}
\usepackage{units}
\usepackage[latin1]{inputenc}
\usepackage{amsfonts}
\usepackage{amssymb}
\usepackage{babel}
\usepackage{color}
\usepackage[dvipsnames]{xcolor} 
\usepackage{cite}
\def\@fmsl@sh#1#2#3{\m@th\ooalign{$\hfil#1\mkern#2/\hfil$\crcr$#1#3$}}
 \def\eq#1\en{\begin{equation}#1\end{equation}}
\def\s[#1,#2]{[#1\stackrel{\star}{,}#2]}
\def\sx[#1,#2]{[#1\stackrel{\star_{x}}{,}#2]}

\newcommand{\logbox}{{\ln\!\left(\frac{\Box}{\mu^2}\right)}}

\newcommand{\GN}{{G_{\rm N}}}
\newcommand{\LP}{{\ell_{\rm P}}}
\newcommand{\lP}{{\ell_{\rm P}}}
\newcommand{\lPs}{{\ell_{\rm P}^2}}
\newcommand{\LPS}{{\ell_{\rm P}^2}}

\newcommand{\Rs}{{R_{\rm s}}}

\usepackage[normalem]{ulem}

\def\gsim{\mathrel{\rlap{\lower4pt\hbox{\hskip1pt$\sim$}}
		\raise1pt\hbox{$>$}}}       

\newcommand{\nc}{\newcommand}
\nc{\beq}{\begin{equation}}
\nc{\eeq}{\end{equation}}
\nc{\beqa}{\begin{eqnarray}}
\nc{\eeqa}{\end{eqnarray}}

\def\bc{\begin{center}}
\def\ec{\end{center}}

\def\to{\rightarrow}

\def\gsim{\mathrel{\mathpalette\atversim>}}

\def\bc{\begin{center}}
\def\ec{\end{center}}

\def\gsim{\mathrel{\rlap{\lower4pt\hbox{\hskip1pt$\sim$}}

    \raise1pt\hbox{$>$}}}       

\def\gsim{\mathrel{\rlap{\lower4pt\hbox{\hskip1pt$\sim$}}
    \raise1pt\hbox{$>$}}}       

\begin{document}
\makeatletter
\def\fmslash{\@ifnextchar[{\fmsl@sh}{\fmsl@sh[0mu]}}
\def\fmsl@sh[#1]#2{%
  \mathchoice
    {\@fmsl@sh\displaystyle{#1}{#2}}%
    {\@fmsl@sh\textstyle{#1}{#2}}%
    {\@fmsl@sh\scriptstyle{#1}{#2}}%
    {\@fmsl@sh\scriptscriptstyle{#1}{#2}}}
\def\@fmsl@sh#1#2#3{\m@th\ooalign{$\hfil#1\mkern#2/\hfil$\crcr$#1#3$}}
\makeatother

\thispagestyle{empty}
\begin{titlepage}
\boldmath
\begin{center}
    {\Large \bf On Modelling the Surfaces of Celestial Bodies in Quantum Gravity}
\end{center}
\unboldmath
\vspace{0.2cm}
\begin{center}
{\large Xavier Calmet}\footnote[1]{\href{x.calmet@sussex.ac.uk}{x.calmet@sussex.ac.uk}}
{\large and} {\large Marco~Sebastianutti}\footnote[2]{\href{m.sebastianutti@sussex.ac.uk}{m.sebastianutti@sussex.ac.uk}}
 \end{center}
\begin{center}
{\sl Department of Physics and Astronomy, 
University of Sussex, Brighton, BN1 9QH, United Kingdom
}
\end{center}
\vspace{5cm}
\begin{abstract}
\noindent
We discuss how to model the surface of celestial bodies (such as stars) in quantum gravity to ensure the regularity of quantum corrections to classical solutions of general relativity at the surface of such bodies. Specifically, we use the Vilkovisky--DeWitt unique effective action to calculate universal quantum corrections to the exterior metric for a class of stellar models. Previous descriptions, obtained via a Heaviside density profile, are ``pathological'' at the surface of the star due to the divergence of the metric functions and associated curvature invariants. Introducing a modified version of the Tolman~{\rm VII} density profile, we determine the minimal degree of differentiability required for this function to generate regular quantum corrections at the star's surface.
\end{abstract}  
\end{titlepage}



\newpage
\renewcommand{\thefootnote}{\arabic{footnote}}

\section{Introduction}
The aim of this paper is to study the modelling of boundaries of celestial bodies in quantum gravity. In general relativity sharp boundaries, if handled properly via the Israel junction conditions~\cite{Israel:1966rt,Poisson:2009pwt}, do not lead to discontinuities in the metrics. In quantum gravity, however, modelling the interior of a matter source with a density profile that is discontinuous at the surface, such as the Heaviside function, leads to divergences in the metric and associated scalar quantities. Focusing for simplicity on the exterior spacetime region and considering static and spherically symmetric sources, we demonstrate how to resolve this issue by adopting smoother profiles for the matter density.

We work within the framework of the Vilkovisky--DeWitt unique effective action~\cite{Barvinsky:1983vpp,Vilkovisky:1984st,Barvinsky:1985an,Barvinsky:1987uw,Barvinsky:1990up,Buchbinder:1992gdx,Calmet:2018elv}, in which quantum gravitational corrections to the metric of a variety of spacetimes have been recently calculated in the literature~\cite{Calmet:2019eof,Calmet:2021lny,Calmet:2021stu,Calmet:2023met,Calmet:2024pve,Antonelli:2025zwc,Perrucci:2024qrr,Cheong:2023oik,Cheong:2025scf}. Adopting this framework implies that these corrections, which are gauge invariant, are universal and apply to any model of quantum gravity that has general relativity as its low energy limit.

Among the many spacetimes analysed in the literature, quantum corrections to the background metrics of a constant density star in stable equilibrium were obtained in~\cite{Calmet:2019eof}. In that work, the authors perturbatively solved the modified Einstein equations that are obtained from the variation of the unique effective action up to second order in curvature. Even though these perturbative solutions are regular far from the star's surface $\Rs$, the quantum-corrected spacetime is still not free of problematic aspects. In fact, since the Heaviside function used to describe the star's constant density profile is discontinuous at the surface of the spherical object, both the interior and exterior quantum corrections diverge at $r=\Rs$. In turn, the surface gravity $\kappa$ and curvature invariants such as the Ricci scalar $R$ and Kretschmann scalar ${\cal K}\equiv R_{\mu\nu\rho\sigma}R^{\mu\nu\rho\sigma}$, which remain everywhere finite at the background level, diverge at the star's surface once the quantum corrections are included, making $r=\Rs$ a truly singular point of the manifold. To circumvent this issue, the metric perturbations are usually considered to be valid up to one Planck length from the surface of the spherical object, i.e.~$\abs{r-\Rs}\geq\lP$. The ``pathological'' behaviour of the spacetime at $r=\Rs$ is then considered unphysical, being merely an artifact of the discontinuity in the Heaviside density profile~\cite{Calmet:2019eof,Perrucci:2024qrr}.

In order to cure the aforementioned divergences, the sharp boundary at the surface of the matter source must be smoothed~\cite{Satz:2004hf,Mazzitelli:2011st,Boasso:2025ofd}. We therefore study a class of stellar models characterized by a density function that smoothly approaches $\Rs$, and examine its impact on the regularity of the metric perturbations at the surface of the spherical object. In particular, inspired by the form of the Tolman {\rm VII} density profile~\cite{Tolman:1939jz}, we consider a function which, depending on the value of its parameter $\lambda$, can interpolate between a Heaviside and a class $C^k$ density profile as $r\to R_{\rm s}^{-}$. On the background spacetime of a static star with such non-uniform density, we calculate, perturbatively at second order in $\lP\equiv\sqrt{\hbar\GN}$ and first order in compactness $C\equiv\frac{2\GN M}{\Rs}$, the quantum corrections to the exterior background metric. We elaborate on the conditions that, through the parameter $\lambda$, must be imposed on the density profile in order to achieve a certain regularity of the exterior solutions at the surface of the spherical object. We first restrict $\lambda$ to values that remove the divergences in the metric as $r\to R_{\rm s}^+$, then we determine the range of $\lambda$ that keeps second-derivative curvature invariants finite at the surface of the matter source. Finally, we identify the minimum value of $\lambda$ required for the spacetime to be of class $C^k$ throughout the region $r\ge\Rs$. 

Furthermore, we study the behaviour of the quantum corrections far away from the surface of the star, i.e.~for $r\gg\Rs$. We observe that, at next-to-leading order in the asymptotic expansion, the exterior metric acquires a dependence on the star's interior composition, a quantum ``hair'' feature which was originally observed in~\cite{Calmet:2021stu}. As an example of physical observable, we take the deflection angle $\delta\phi$ and evaluate the contribution of our quantum corrections to its background value~\cite{Wald:1984rg}. Since such corrections turn out to be $\lPs/R_{\rm s}^2$-suppressed for a star in stable equilibrium, the deviation of $\delta\phi$ from its classical value is not measurable in practice. Nonetheless, our derivation demonstrates that quantum gravitational corrections to observable quantities can be analytically calculated in a model independent way and from first principles.

This paper is organised as follows: in Sec.~\ref{sec:intro} we derive the modified Einstein equations starting from the unique effective action. In Sec.~\ref{sec:modifiedTolman}, we elaborate on the reason why the quantum-corrected solutions at the surface of a static and constant density star diverge, which, in turn, motivate the introduction of a modified version of the Tolman {\rm VII} density profile. In Sec.~\ref{sec:solutions}, we present the exterior perturbative solutions on the background spacetime of the previous section, focusing on the behaviour of the metric close to the surface of the spherical object and very far away from it. In Sec.~\ref{sec:conclusions}, we discuss the results and draw our conclusions.

\section{Unique effective action framework}\label{sec:intro}
Quantum gravitational corrections to classical solutions of general relativity can be found by solving the field equations derived from the variation of the Vilkovisky--DeWitt unique effective action~\cite{Barvinsky:1983vpp,Vilkovisky:1984st,Barvinsky:1985an,Barvinsky:1987uw,Barvinsky:1990up,Buchbinder:1992gdx,Calmet:2018elv}. These corrections are model independent and apply to any ultraviolet (UV) complete theory of quantum gravity that recovers general relativity at low energies. The unique effective action, truncated at second order in curvature, is obtained by integrating out massless field fluctuations and is given by:
\begin{equation}\label{eq:UEA}
    \Gamma = \Gamma_{\rm L} + \Gamma_{\rm NL} +\Gamma_{\rm m},
\end{equation}
with a local part
\begin{align}
    \Gamma_{\rm L} = \int d^4x\,\sqrt{|g|} \bigg[& \frac{M_{\rm P}^2 }{2}R + c_1(\mu)R^2 + c_2(\mu)R_{\mu\nu}R^{\mu\nu}\nonumber\\
    &+ c_3(\mu)R_{\mu\nu\rho\sigma}R^{\mu\nu\rho\sigma} + \order{M_{\rm P}^{-2}} \bigg],
\end{align}
a non-local part 
\begin{align}
    \Gamma_{\rm NL}= -\int d^4x\,\sqrt{|g|} \bigg[& \alpha R\,\logbox R
    +\beta R_{\mu\nu}\, \logbox R^{\mu\nu}\nonumber\\
    &+\gamma R_{\mu\nu\rho\sigma} \,\logbox R^{\mu\nu\rho\sigma} + \order{M_{\rm P}^{-2}} \bigg],
\end{align}
and with the matter sector enclosed in $\Gamma_{\rm m}$. In this work, $M_{\rm P}\equiv\sqrt{\hbar/8\pi\GN}$ is the reduced Planck mass, $\mu$ is a renormalization scale~\cite{Buchbinder:1992gdx,Lombardo:1996gp} and $c_1(\mu)$, $c_2(\mu)$ and $c_3(\mu)$ are local Wilson coefficients whose values cannot be predicted without a UV complete theory of quantum gravity. For a string theory example, where the expressions of the local Wilson coefficients are computed explicitly, see~\cite{Calmet:2024neu}. Furthermore, $\alpha$, $\beta$ and $\gamma$ are the non-local Wilson coefficients whose known values depend on the spin of the massless fluctuating field that has been integrated out. For their numerical values, we refer the reader to Tab.~1 of~\cite{Calmet:2019eof} and references therein.

By varying the above action with respect to the metric, we obtain the quantum-corrected Einstein equations at second order in curvature:
\begin{equation}\label{eq:EOMs}
    G_{\mu\nu}+16\pi\hbar\GN\left(H_{\mu\nu}^{\rm L}+H_{\mu\nu}^{\rm NL}\right)=8\pi\GN T_{\mu\nu},
\end{equation}
where the local and non-local additions are given respectively by:
\begin{align}\label{eq:HL}
    H_{\mu\nu}^{\rm L}=&\,\,\bar{c}_1(\mu)\left(2RR_{\mu\nu}-\frac{1}{2}g_{\mu\nu}R^2+2g_{\mu\nu}\Box R-2\nabla_\mu\nabla_\nu R\right)\notag\\
    &+\bar{c}_2(\mu)\left(2\tensor{R}{^\rho_\mu}\tensor{R}{_{\nu\rho}}-\frac{1}{2}\tensor{g}{_{\mu\nu}}\tensor{R}{^\rho_\sigma}\tensor{R}{^\sigma_\rho}+\Box \tensor{R}{_{\mu\nu}}+\frac{1}{2}\tensor{g}{_{\mu\nu}}\Box \tensor{R}{}-2\tensor{\nabla}{^\rho}\tensor{\nabla}{_{(\mu}} \tensor{R}{_{\nu)\rho}}\right),
\end{align}
and 
\begin{align}\label{eq:HNL}
    H_{\mu\nu}^{\rm NL}=&\,-\bar{\alpha}\bigg(2R_{\mu\nu}-\frac{1}{2}g_{\mu\nu}R+2g_{\mu\nu}\Box-2\nabla_\mu\nabla_\nu\bigg)\logbox R\notag\\
    &\,-\bar{\beta}\left(2\tensor{\delta}{^\sigma_{(\mu}}\tensor{R}{_{\nu)\rho}}-\frac{1}{2}\tensor{g}{_{\mu\nu}}\tensor{R}{^\sigma_\rho}+\tensor{\delta}{^\sigma_\nu}\tensor{g}{_{\mu\rho}}\Box+\tensor{g}{_{\mu\nu}}\tensor{\nabla}{^\sigma}\tensor{\nabla}{_\rho}-2\tensor{\delta}{^\sigma_{(\mu}}\tensor{\nabla}{_{|\rho|}}\tensor{\nabla}{_{\nu)}}\right)\!\logbox\tensor{R}{^\rho_\sigma},
\end{align}
with the re-defined coefficients $\bar{c}_1(\mu)=c_1(\mu)-c_3(\mu)$, $\bar{c}_2(\mu)=c_2(\mu)+4c_3(\mu)$ and $\bar{\alpha}=\alpha-\gamma$, $\bar{\beta}=\beta+4\gamma$ that originate from the local and non-local Gauss--Bonnet identities~\cite{Calmet:2018elv}. These identities are applied to the effective action~\eqref{eq:UEA} to remove the Riemann tensor terms in favour of the Ricci scalar and Ricci tensor ones.

Solutions to the equations of motion~\eqref{eq:EOMs} can be found perturbatively in the Planck length~$\LP\equiv \sqrt{\hbar\GN}$. By perturbing the metric as
\begin{equation}\label{eq:gmunu}
    g_{\mu\nu}=\bar{g}_{\mu\nu}+h_{\mu\nu},
\end{equation}
where $\bar{g}_{\mu\nu}$ is the classical background and $h_{\mu\nu}$ the quantum perturbation, the perturbed Einstein equations, up to second order in $\lP$, read:
\begin{equation}\label{eq:EOMsLP}
    \delta G_{\mu\nu}+16\pi\LPS\left(\bar{H}_{\mu\nu}^{\rm L}+\bar{H}_{\mu\nu}^{\rm NL}\right)=0.
\end{equation}
The linearized Einstein tensor is given by:
\begin{equation}
    2\delta G_{\mu\nu}=-\tensor{\bar{R}}{}\tensor{h}{_{\mu\nu}}+\tensor{\bar{g}}{_{\mu\nu}}\tensor{\bar{R}}{_{\rho\sigma}}\tensor{h}{^{\rho\sigma}}+2\tensor{\bar{\nabla}}{_\rho}\tensor{\bar{\nabla}}{_{(\mu}}\tensor{h}{_{\nu)}^{\rho}}-\tensor{\bar{g}}{_{\mu\nu}}\tensor{\bar{\nabla}}{_{\rho}}\tensor{\bar{\nabla}}{_{\sigma}}\tensor{h}{^{\rho\sigma}}-\tensor{\bar{\nabla}}{_{\mu}}\tensor{\bar{\nabla}}{_{\nu}}\tensor{h}{}-\bar{\Box}\tensor{h}{_{\mu\nu}}+\tensor{\bar{g}}{_{\mu\nu}}\bar{\Box}\tensor{h}{},
\end{equation}
with $h\equiv \tensor{h}{^\rho_\rho}$, and $\bar{H}_{\mu\nu}^{\rm L}$ and $\bar{H}_{\mu\nu}^{\rm NL}$ the background-evaluated~\eqref{eq:HL} and~\eqref{eq:HNL}, respectively.\footnote{With the exception of the non-local Wilson coefficients $\bar{\alpha}$ and $\bar{\beta}$, we denote with an over-bar tensorial quantities that are evaluated on the background spacetime.} Once the classical background metric is fixed and an ansatz for the perturbation~$h_{\mu\nu}$ is provided, the $\lPs$-corrected Einstein equations~\eqref{eq:EOMsLP} can be solved perturbatively. Herein, we take the background spacetime of a static and spherically symmetric matter source as modelled by a perfect fluid with non-uniform density. The metric perturbation is also assumed static and spherically symmetric.

In the following sections, starting from the constant density case, we elaborate on the reason why quantum corrections evaluated at the surface of the matter source diverge. Next, we replace the Heaviside function modelling the density profile of the spherical object with a function that is smoother as $r\to R_{\rm s}^-$. Exterior perturbative solutions for this new density profile are later found in Sec.~\ref{sec:solutions}.

\section{Modified Tolman {\rm VII} density profile}\label{sec:modifiedTolman}
The quantum corrections to the classical metric of a static, spherically symmetric and constant-density matter source (modelling e.g.~a star in stable equilibrium) were computed in~\cite{Calmet:2019eof} and earlier in~\cite{Satz:2004hf}, using the density profile:
\begin{equation}\label{eq:STEP}
    \rho=
        \begin{dcases} 
            \rho_0 & r < \Rs \\
            0 & r\geq \Rs
        \end{dcases},
\end{equation}
where $\rho_0=\frac{3M}{4\pi R_{\rm s}^3}$, $M$ being the total mass of the spherical object and $\Rs$ its surface radius. The background spacetime associated with this configuration is described by two metrics:~interior Schwarzschild for~$0 \leq r\leq \Rs$
\begin{align}
    ds^2&\equiv \bar{g}_{\mu\nu}^{\,\rm int}dx^\mu dx^\nu\nonumber\\
    &=\frac{1}{4}\left(3\sqrt{1-\frac{2G_{\rm N}M}{R_{\rm s}}}-\sqrt{1-\frac{2G_{\rm N}Mr^2}{R_{\rm s}^3}}\right)^2 dt^2-\left(1-\frac{2G_{\rm N}Mr^2}{R_{\rm s}^3}\right)^{-1} dr^2-r^2d\Omega^2,
\end{align}
and (exterior) Schwarzschild for~$r\geq \Rs$
\begin{equation}\label{eq:Schw}
    ds^2\equiv \bar{g}_{\mu\nu}^{\,\rm ext}dx^\mu dx^\nu=\left(1-\frac{2G_{\rm N}M}{r}\right) dt^2-\left(1-\frac{2G_{\rm N}M}{r}\right)^{-1} dr^2-r^2d\Omega^2.
\end{equation}
Even though the two classical backgrounds are continuous at $r=\Rs$, the corresponding quantum corrections computed on these spacetimes, as well as any curvature invariants built from them, diverge at the junction between the interior and exterior regions.
Quantum corrections at order $\lPs$ are obtained via a perturbative expansion in the parameter $C\equiv\frac{2\GN M}{\Rs}$, the star's compactness, and, for the exterior spacetime region, are given by~\cite{Calmet:2019eof}:
\begin{align}
    h_{tt}^{\rm ext}&=96\pi\left(\bar{\alpha}+\bar{\beta}\right)\frac{\LPS\!\cdot\! C}{R_{\rm s}^2}\left(2\frac{\Rs}{r}+\ln{\frac{r-\Rs}{r+\Rs}}\right)+\frac{\LPS}{R_{\rm s}^2}\!\cdot\!\order{C^2},\label{eq:h00STEP}\\[0.5em]
    h_{rr}^{\rm ext}&=192\pi\bar{\alpha}\frac{\LPS\!\cdot\! C}{R_{\rm s}^2}\frac{R_{\rm s}^3}{r(r^2-R_{\rm s}^2)}+\frac{\LPS}{R_{\rm s}^2}\!\cdot\!\order{C^2}.\label{eq:h11STEP}
\end{align}
As was first pointed out in~\cite{Satz:2004hf} and later in~\cite{Mazzitelli:2011st}, the cause of this ``pathological'' behaviour lies in the discontinuity at the surface of the density profile~\eqref{eq:STEP}. To resolve this issue a smoother density function is required. In this work, focusing on the exterior spacetime region, whose classical background is given by the Schwarzschild metric~\eqref{eq:Schw}, we determine the degree of regularity that a $\rho(r)$ of class $C^k$ induces on the quantum corrections $h_{tt}^{\rm ext}$ and $h_{rr}^{\rm ext}$.

\begin{figure}[ht!]
\centering 
\includegraphics[width=0.575\textwidth]{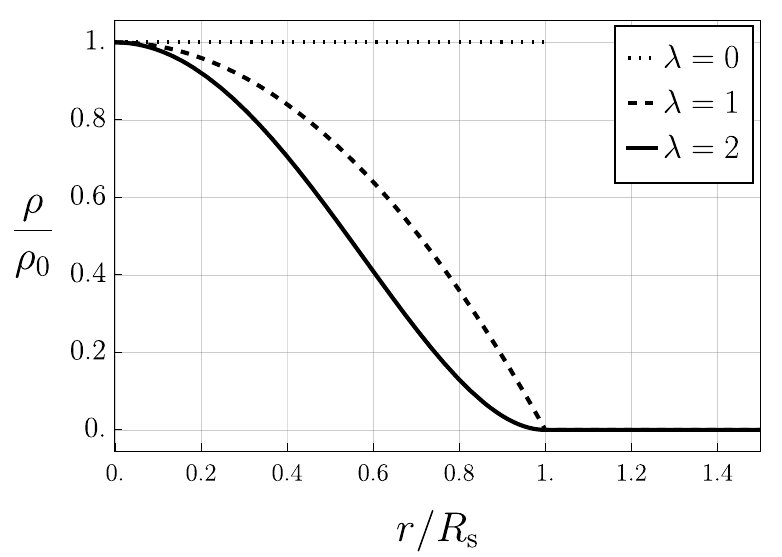}
\caption{Modified Tolman {\rm VII} density profile (in units of $\rho_0$) for different values of $\lambda$. Any $\lambda>0$ produces a continuous density profile at $r=\Rs$; $\lambda=1$ corresponds to the original Tolman {\rm VII} density profile, any $\lambda>2$ corresponds to a $\rho(r)$ of class at least $C^2$.}
\label{fig:MTVII}
\end{figure}

To this end, we introduce a modified version of the Tolman {\rm VII} density profile~\cite{Tolman:1939jz}:
\begin{equation}\label{eq:MTVII}
    \rho=\begin{dcases} 
                \rho_0\left(1-\frac{r^2}{R_{\rm s}^2}\right)^\lambda & r < \Rs \\
                0 & r \geq \Rs
            \end{dcases},
\end{equation}
where $\lambda$, the parameter that controls the smoothness of the density profile at the surface, is restricted to $\mathbb{R}_{\ge0}$ to prevent $\rho$ from increasing with radius and diverging at~$r=\Rs$. Except $\lambda=0$, for which the above reduces to~\eqref{eq:STEP}, all other $\lambda$ values produce a continuous and decreasing function of $r$; $\lambda=1$ corresponding to the original Tolman {\rm VII} density profile, see Fig.~\ref{fig:MTVII}. Unlike the constant density case, for which $\rho_0=\frac{3M}{4\pi R_{\rm s}^3}$, now $\rho_0\equiv\rho(0)$ stands for the central density of the matter source, whose value depends on~$\lambda$. In fact, the Arnowitt--Deser--Misner ({\rm ADM}) mass of the system~\cite{Arnowitt:1962hi} for a static star~\cite{Friedman:2013xza} reads:
\begin{equation}
    M\equiv M_{\rm ADM}=4\pi \int_{0}^{\infty}\rho(r)r^2dr=2{\rm B}\!\left(\tfrac{3}{2},1+\lambda\right)\!\pi R_{\rm s}^3 \rho_0,
\end{equation}
where ${\rm B}\!\left(a,b\right)\equiv\frac{\Gamma\left( a\right)\Gamma\left( b\right)}{\Gamma\left(a+b\right)}$ denotes the beta function, $\Gamma(a)$ the gamma function, $a$ and $b$ being complex numbers. From the above, it is immediate to find
\begin{equation}
    \rho_0=\frac{3M}{4\pi R_{\rm s}^3}\frac{2/3}{{\rm B}\!\left(\frac{3}{2},1+\lambda\right)},
\end{equation}
a monotonically increasing function of $\lambda$. For any $\lambda>0$ one has $\rho_0>\frac{3M}{4\pi R_{\rm s}^3}$, which means that, for a decreasing $\rho(r)$, a higher central density than in the $\lambda = 0$ case is required for $M$ to remain the {\rm ADM} mass of the system.

In order to obtain the interior background metric originating from the density profile~\eqref{eq:MTVII}, one is required to solve the corresponding Tolman--Oppenheimer--Volkoff (TOV) system of coupled differential equations~\cite{Wald:1984rg}. Finding analytic solutions for this $\rho(r)$ is feasible only for specific values of $\lambda$, e.g.~$\lambda=1$ corresponds to the original Tolman {\rm VII} metric~\cite{Tolman:1939jz}, and lies beyond the scope of the present manuscript.

In the next section, we derive the exterior quantum corrections by solving perturbatively at first order in compactness the $\lPs$-corrected Einstein equations~\eqref{eq:EOMsLP}. In doing this, we replace the constant density profile~\eqref{eq:STEP} with the modified Tolman {\rm VII} defined in~\eqref{eq:MTVII}. Then, we study the solution $h_{\mu\nu}^{\rm ext}$ in two limits: as it approaches the star's surface ($r\to R_{\rm s}^+$) and far away from it ($r\gg\Rs$).

\section{Exterior perturbative solutions}\label{sec:solutions}
We now focus on the spacetime region outside the spherical object, i.e.~$r\geq \Rs$, whose background is given by the Schwarzschild metric~\eqref{eq:Schw}, and follow~\cite{Calmet:2019eof} to determine the quantum corrections up to second order in $\ell_{\rm P}$, which, due to the non-local nature of the operator~$\logbox$, turn out to depend on the interior composition of the matter source. In order to find the metric perturbation $h_{\mu\nu}^{\rm ext}$, we evaluate the quantum-corrected Einstein equations~\eqref{eq:EOMsLP} on the Schwarzschild background and expand them to linear order in the compactness parameter, neglecting terms of order $\order{C^2}$. We remark that, while the local term $\bar{H}_{\mu\nu}^{\rm L}$ vanishes exactly in vacuum, $\bar{H}_{\mu\nu}^{\rm NL}$ contains the logbox operator, whose non-local action on the background-evaluated Ricci scalar $\bar{R}$ and Ricci tensor $\tensor{\bar{R}}{_{\mu\nu}}$ of Eq.~\eqref{eq:HNL} produces a non-vanishing result. In fact, by expressing these curvature tensors in terms of the matter sector:
\begin{align}
    \bar{R}&=8\pi\GN\left(3p-\rho\right),\\[0.25em]
    \tensor{\bar{R}}{_{\mu\nu}}&=8\pi\GN\left[\left(\rho+p\right)\bar{u}_\mu \bar{u}_\nu+\frac{1}{2}\left(p-\rho\right)\tensor{\bar{g}}{_{\mu\nu}}\right],
\end{align}
where
\begin{equation}
    \bar{u}^{\mu}=\left(\frac{1}{\sqrt{\bar{g}_{tt}}},0,0,0\right),
\end{equation}
and $p=p(r)$ is the pressure of the perfect fluid, we can expand $\logbox\rho$ as done in~\cite{Calmet:2019eof}, with the density profile that is now the one of Eq.~\eqref{eq:MTVII}. Despite the fact that $\rho=0$ outside the matter source, the non-local action of the logbox operator on the density profile generates a non-vanishing function of $r$ even for $r>\Rs$. In turn, this results in a non-vanishing $\bar{H}_{\mu\nu}^{\rm NL}$, serving as the inhomogeneous term in the $\LPS$-corrected Einstein equations~\eqref{eq:EOMsLP}, which can be solved perturbatively in the compactness parameter $C$.\footnote{The contribution of $\logbox p$ to the background-evaluated $H_{\mu\nu}^{\rm NL}$ of Eq.~\eqref{eq:HNL} can be neglected without explicitly acting with the logbox operator on $p$. In fact, since $p\sim\order{C}$, the function of $r$ resulting from the expansion of the non-local operator on the pressure contributes to $\bar{H}_{\mu\nu}^{\rm NL}$ quadratically in~$C$. As our perturbative expansion is truncated at linear order in the compactness parameter, we can safely disregard any term containing $\logbox p$ from the outset.}

Imposing the metric perturbation $h^{\rm ext}_{\mu\nu}$ to be static and spherically symmetric, the resulting exterior quantum corrections to linear order in $C$ read:
\begin{align}
    h_{tt}^{\rm ext}=&-64\pi\left(\bar{\alpha}+\bar{\beta}\right)\frac{\lPs\!\cdot\! C}{R_{\rm s}^2}{}_2F_1\!\left(1,\frac{3}{2};\frac{5}{2}+\lambda;\frac{R_{\rm s}^2}{r^2}\right)\frac{R_{\rm s}^3}{r^3}+K_2+\frac{K_1}{r}-K_2\frac{\Rs\!\cdot\!C}{r}+\frac{\LPS}{R_{\rm s}^2}\!\cdot\!\order{C^2},\label{eq:h00MTVII}\\
    h_{rr}^{\rm ext}=&\;192\pi\bar{\alpha}\frac{\lPs\!\cdot\! C}{R_{\rm s}^2}{}_2F_1\!\left(1,\frac{5}{2};\frac{5}{2}+\lambda;\frac{R_{\rm s}^2}{r^2}\right)\frac{R_{\rm s}^3}{r^3}+\frac{K_1}{r}+2\frac{K_1\Rs\!\cdot\! C}{r^2}+\frac{\lPs}{R_{\rm s}^2}\!\cdot\!\order{C^2},\qquad\qquad\qquad\label{eq:h11MTVII}
\end{align}
where ${}_2F_1\left(a,b;c;z\right)$ denotes the Gauss's hypergeometric function, with $a$, $b$ and $c$ complex parameteres and $z$ a complex variable.
The integration constants $K_1$ and $K_2$ arise from solving the equations of motion~\eqref{eq:EOMsLP} perturbatively in compactness.
To preserve the asymptotic flatness of the background spacetime, $K_2$ must vanish; additionally, $K_1$ is set to zero to ensure consistency with the classical Newtonian limit.
Note that the constant density solutions of Eqs.~\eqref{eq:h00STEP} and~\eqref{eq:h11STEP} are recovered, as expected, for $\lambda=0$. 

Once we have the perturbed metric $g_{\mu\nu}^{\rm ext}=\bar{g}_{\mu\nu}^{\,\rm ext}+h_{\mu\nu}^{\rm ext}$, it is immediate to obtain scalar quantities such as the surface gravity $\kappa$ (see e.g.~\cite{Wald:1984rg} for its definition) and the Ricci and Kretschmann scalars, $R$ and ${\cal K}$ respectively. At lowest order in the expansion parameter $C$, always for $r\geq\Rs$, the former is given by:
\begin{equation}\label{eq:kappa}
    \kappa=\frac{\Rs\!\cdot\! C}{2r^2}+96\pi\left(\bar{\alpha}+\bar{\beta}\right)\frac{\lPs\!\cdot\! C}{R_{\rm s}^3}{}_2F_1\!\left(1,\frac{5}{2};\frac{5}{2}+\lambda;\frac{R_{\rm s}^2}{r^2}\right)\frac{R_{\rm s}^4}{r^4}+\frac{\lPs}{R_{\rm s}^3}\!\cdot\!\order{C^2},
\end{equation}
while the second-derivative curvature invariants read, respectively:
\begin{align}
    R&=384\pi\left(3\bar{\alpha}+\bar{\beta}\right)\frac{\LPS\!\cdot\! C}{R_{\rm s}^4}{}_2F_1\!\left(2,\frac{5}{2};\frac{5}{2}+\lambda;\frac{R_{\rm s}^2}{r^2}\right)\frac{R_{\rm s}^5}{r^5}+\frac{\LPS}{R_{\rm s}^4}\!\cdot\!\order{C^2},\\[0.5em]
    {\cal K}&=\frac{12R_{\rm s}^2\!\cdot\!C^2}{r^6}+3840\pi\bar{\beta}\frac{\LPS\!\cdot\! C^2}{R_{\rm s}^6}{}_2F_1\!\left(1,\frac{7}{2};\frac{5}{2}+\lambda;\frac{R_{\rm s}^2}{r^2}\right)\frac{R_{\rm s}^8}{r^8}+\frac{\LPS}{R_{\rm s}^6}\!\cdot\!\order{C^3}.
\end{align}
In particular, $R\neq0$ implies that the exterior region is not in vacuum once the first order quantum corrections are added to the Schwarzschild background. This means that, outside the matter source, a non-vanishing effective stress-energy tensor is generated at the quantum level, whose effective energy density and pressure components decrease monotonically with increasing distance from the source.

In the following, we focus on the two limiting behaviours of our quantum-corrected metric: at the star's surface ($r\to R_{\rm s}^+$) and far away from it ($r\gg\Rs$).

\subsection{Quantum corrections at the surface}
Depending on the value of $\lambda$, the metric perturbations $h_{tt}^{\rm ext}$ and $h_{rr}^{\rm ext}$ may or may not be divergent at $r=\Rs$. If the density profile~\eqref{eq:MTVII} is continuous, i.e. if $\lambda>0$, then $\logbox\rho$ is not divergent at the surface of the matter source. However, to obtain perturbative solutions that are finite at the surface, also the first derivative of $\logbox\rho$ is required to remain finite at $r=\Rs$. This condition is satisfied in case $\rho(r)$ is a function of class at least $C^1$, which, for the modified Tolman {\rm VII} density profile, corresponds to having $\lambda>1$. In this case, both $h_{tt}^{\rm ext}$ and $h_{rr}^{\rm ext}$ remain finite at the surface of the spherical object; as can be observed by taking the limit $r\to R_{\rm s}^+$, for which Eqs.~\eqref{eq:h00MTVII} and~\eqref{eq:h11MTVII} reduce to the following expressions:
\begin{align}
    \lim_{r\to R_{\rm s}^+}{h_{tt}^{\rm ext}}&=-64\pi\left(\bar{\alpha}+\bar{\beta}\right)\frac{\LPS\!\cdot \!C}{R_{\rm s}^2}\frac{3/2+\lambda}{\lambda}+\frac{\LPS}{R_{\rm s}^2}\!\cdot\!\order{C^2},&&\lambda\ge0,\label{eq:h00MTVIIx1}\\[0.25em]
    \lim_{r\to R_{\rm s}^+}{h_{rr}^{\rm ext}}&=192\pi\bar{\alpha}\frac{\LPS\!\cdot\! C}{R_{\rm s}^2}\frac{3/2+\lambda}{\lambda-1}+\frac{\LPS}{R_{\rm s}^2}\!\cdot\!\order{C^2},&&\lambda\ge1.\label{eq:h11MTVIIx1}
\end{align}
Note that the above (and below) limits are valid only for the specified ranges of $\lambda$. If $\lambda>1$, the surface gravity~\eqref{eq:kappa} is also finite as $r\to R_{\rm s}^+$, as can be seen from:
\begin{equation}
    \lim_{r\to R_{\rm s}^+}{\kappa}=\frac{C}{2\Rs}+96\pi\left(\bar{\alpha}+\bar{\beta}\right)\frac{\LPS\!\cdot \!C}{R_{\rm s}^3}\frac{3/2+\lambda}{\lambda-1}+\frac{\LPS}{R_{\rm s}^3}\!\cdot\!\order{C^2},\qquad \lambda\ge1.
\end{equation}
Even though the quantum-corrected metric $g_{\mu\nu}^{\rm ext}$ is regular at the surface, curvature invariants such as the Ricci scalar $R$ and the Kretschmann scalar ${\cal K}$ may still diverge at $r=\Rs$. To avoid this from happening, the second derivative of $h_{tt}^{\rm ext}$ and the first of $h_{rr}^{\rm ext}$ need to remain finite at this point. Such a condition requires an even smoother density profile, specifically a $\rho(r)$ of class at least $C^2$, which, for the density function in Eq.~\eqref{eq:MTVII}, implies $\lambda>2$. This can be verified by evaluating the Ricci and Kretschmann scalars in the limit $r\to R_{\rm s}^+$:
\begin{align}
    \lim_{r\to R_{\rm s}^+}{R}&=384\pi\left(3\bar{\alpha}+\bar{\beta}\right)\frac{\LPS\!\cdot\! C}{R_{\rm s}^4}\frac{3/2+\lambda}{\lambda-2}\frac{1/2+\lambda}{\lambda-1}+\frac{\LPS}{R_{\rm s}^4}\!\cdot\!\order{C^2}, && \lambda\ge2,\label{eq:R}\\[0.25em]
    \lim_{r\to R_{\rm s}^+}{\cal K}&=\frac{12C^2}{R_{\rm s}^4}+3840\pi\bar{\beta}\frac{\LPS\!\cdot\!C^2}{R_{\rm s}^6}\frac{3/2+\lambda}{\lambda-2}+\frac{\LPS}{R_{\rm s}^6}\!\cdot\!\order{C^3}, && \lambda\ge2.\label{eq:K}
\end{align}
More in general, in order to have a $g_{\mu\nu}^{\rm ext}$ of class $C^k$ at the star's surface, the modified Tolman~{\rm VII} density profile must be at least of class $C^{k+1}$, which corresponds to~$\lambda>k+1$.

We remark that the above limits, following from the hypergeometric identity~\cite{Duverney:2024qzj}:
\begin{equation}\label{eq:hyper}
    {}_2F_1\!\left(a,b;c;1\right)\equiv\frac{\Gamma\left(c\right)\Gamma\left(c-a-b\right)}{\Gamma\left(c-a\right)\Gamma\left(c-b\right)}, \qquad \Re\!\left(c-a-b\right)>0,
\end{equation}
are valid only for the specified ranges of $\lambda$. Outside these intervals, all the above expressions fail to reproduce the correct (divergent) behaviours at $r=\Rs$.

\subsection{Quantum corrections far from the surface}
Far away from the spherical object, i.e.~for $r\gg\Rs$, the quantum-corrections~\eqref{eq:h00MTVII} and~\eqref{eq:h11MTVII} can be expanded as:
\begin{align}
    h_{tt}^{\rm ext}&=-64\pi\left(\bar{\alpha}+\bar{\beta}\right)\frac{\LPS\!\cdot\!C}{R_{\rm s}^2}\frac{R_{\rm s}^3}{r^3}\sum_{i=0}^{n}\frac{\left(3/2\right)_i}{\left(5/2+\lambda\right)_i}\left(\frac{\Rs}{r}\right)^{2i}+\frac{\LPS}{R_{\rm s}^2}\!\cdot\!\order{C^2},\label{eq:h00MTVIIxinf}\\[0.5em]
    h_{rr}^{\rm ext}&=192\pi\bar{\alpha}\frac{\LPS\!\cdot\!C}{R_{\rm s}^2}\frac{R_{\rm s}^3}{r^3}\sum_{i=0}^{n}\frac{\left(5/2\right)_i}{\left(5/2+\lambda\right)_i}\left(\frac{\Rs}{r}\right)^{2i}+\frac{\LPS}{R_{\rm s}^2}\!\cdot\!\order{C^2},\label{eq:h11MTVIIxinf}
\end{align}
where, for a fixed $N$, $(z)_N\equiv z(z+1)\dots(z+N-1)=\Gamma\!\left(z+N\right)\!/\Gamma\!\left(z\right)$ denotes the Pochhammer symbol. Both components go as $r^{-3}$ at leading order, 
with the density parameter $\lambda$ entering at next-to-leading order in the series, namely $r^{-5}$. This makes manifest the dependence of the exterior metric on the internal structure of the matter source; a quantum ``hair'' feature already observed in~\cite{Calmet:2021stu} for the case of two nested dust balls, and more recently in~\cite{Perrucci:2024qrr,Cheong:2025scf} for gravastars and spherical objects composed of dust shells, respectively.~Clearly, by virtue of Birkhoff's theorem, this dependence is absent at the classical level.

With the metric perturbations in the above form, we now show how quantum corrections to observable quantities can be computed analytically. An illustrative example is the deviation from straight-line motion experienced by a light ray due to the gravitational field of a star~\cite{Wald:1984rg}; deviation that is encoded in the deflection angle $\delta \phi$. To calculate this quantity, we consider null geodesic trajectories in the perturbed spacetime $g_{\mu\nu}^{\rm ext}=\bar{g}_{\mu\nu}^{\,\rm ext}+h_{\mu\nu}^{\rm ext}$, whose background is the linearized version of Eq.~\eqref{eq:Schw} and whose first order perturbations are given by Eqs.~\eqref{eq:h00MTVIIxinf} and~\eqref{eq:h11MTVIIxinf}. Working on the equatorial plane ($\theta=\pi/2$), the geodesic equation reads:
\begin{equation}\label{eq:rdot}
    \dot{r}^2=-\frac{1}{g_{rr}^{\rm ext}}\left(\frac{E^2}{g_{tt}^{\rm ext}}-\frac{L^2}{r^2}\right),
\end{equation}
where the dot denotes differentiation w.r.t. an affine parameter, $E$ and $L$ being the conserved ``energy'' and ``angular momentum'' of the photon, respectively. Then, the following relation can be straightforwardly obtained:
\begin{equation}\label{eq:Deltaphi}
    \Delta \phi=\int_{r_0}^{\infty}\frac{dr}{r\sqrt{\frac{1}{g_{rr}^{\rm ext}}\left(1-\frac{r^2}{b^2}\frac{1}{g_{tt}^{\rm ext}}\right)}},
\end{equation}
where the impact parameter $b\equiv L/E$ can be written in terms of the turning point (or perihelion) $r_0$ of the trajectory using Eq.~\eqref{eq:rdot}:
\begin{equation}\label{eq:b}
    b=\frac{r_0}{\sqrt{g_{tt}^{\rm ext}|_{r=r_0}}}.
\end{equation}
By expanding its integrand at linear order in $C$, the integral~\eqref{eq:Deltaphi} can be solved by exchanging the order of the series in Eqs.~\eqref{eq:h00MTVIIxinf} and~\eqref{eq:h11MTVIIxinf} with the integration over $r$. After the integral is performed, re-summing the series leads to the below expression for the deflection angle $\delta\phi\equiv\Delta\phi-\pi$:
\begin{equation}
    \delta\phi=\frac{2C\Rs}{b}+128\pi\bar{\beta}\frac{R_{\rm s}\lPs\!\cdot\! C}{b^3}{}_2F_1\!\left(1,2;\frac{5}{2}+\lambda;\frac{R_{\rm s}^2}{b^2}\right)+\frac{\Rs\LPS}{b^3}\!\cdot\!\order{C^2},
\end{equation}
where, since our result is found at first order in $C$, we have replaced $r_0$ with the impact parameter $b$ from Eq.~\eqref{eq:b}. For a light ray that grazes the surface of the spherical object, where the deflection is maximum, using the hypergeometric identity~\eqref{eq:hyper}, the above equation can be written as:
\begin{equation}
    \lim_{b\to R_{\rm s}^+}{\delta\phi}=2C+128\pi\bar{\beta}\frac{\lPs\!\cdot\! C}{R_{\rm s}^2}\frac{\lambda+3/2}{\lambda-1/2}+\frac{\LPS}{R_{\rm s}^2}\!\cdot\!\order{C^2},\qquad \lambda\ge1/2,
\end{equation}
which highlights the monotonically decreasing behaviour of $\delta\phi$ as $\lambda$ increases. In this case, with the turning point approaching the surface of the matter source, i.e.~in the limit $r_0\to R_{\rm s}^+$ (which is equivalent to $b\to R_{\rm s}^+$ at first order in $C$), one must restrict to values of $\lambda>1/2$ in order to prevent $\delta\phi$ from diverging at this point. Additionally, since the above quantum correction is suppressed by a factor of $\lPs/R_{\rm s}^2$, we remark that its contribution remains, for the case here examined, practically unobservable.

\section{Conclusions}\label{sec:conclusions}
In this paper, within the framework of the Vilkovisky--DeWitt unique effective action~\cite{Barvinsky:1983vpp,Vilkovisky:1984st,Barvinsky:1985an,Barvinsky:1987uw,Barvinsky:1990up,Buchbinder:1992gdx,Calmet:2018elv}, we have calculated universal quantum gravitational corrections to the exterior spacetime of a class of stellar models in stable equilibrium, whose interior is described by a modified version of the Tolman~{\rm VII} density profile~\cite{Tolman:1939jz}. We emphasise that these quantum gravitational corrections are model independent, applying to any UV complete theory of quantum gravity that recovers general relativity at low energies. Following what was previously done for a static and constant density star~\cite{Calmet:2019eof}, our results are obtained by perturbatively solving the $\lPs$-corrected Einstein equations~\eqref{eq:EOMsLP} at first order in the compactness parameter $C$, see Eqs.~\eqref{eq:h00MTVII} and~\eqref{eq:h11MTVII}. By adopting the modified Tolman {\rm VII} density profile~\eqref{eq:MTVII}, we have provided an explicit example of how unphysical divergences in $h_{tt}^{\rm ext}$ and $h_{rr}^{\rm ext}$, due to the sharp boundary between the interior and exterior background spacetimes, can be removed. As pointed out in~\cite{Satz:2004hf,Mazzitelli:2011st,Boasso:2025ofd}, this is done by smoothing the Heaviside density profile~\eqref{eq:STEP} as $r\to R_{\rm s}^-$, see Fig.~\ref{fig:MTVII}. In fact, depending on the value of $\lambda$, the parameter in Eq.~\eqref{eq:MTVII} that controls the smoothness of the density function at the star's surface, it is possible to obtain solutions for the exterior quantum corrections that are free of divergences as $r\to R_{\rm s}^+$.

In particular, we have found that a $\rho(r)$ of class at least $C^1$, which for the density profile here considered corresponds to $\lambda>1$, is sufficient to get finite $h_{tt}^{\rm ext}$ and $h_{rr}^{\rm ext}$ as the star's surface is approached, see Eqs.~\eqref{eq:h00MTVIIx1} and~\eqref{eq:h11MTVIIx1}. With the above restriction on $\lambda$, also the surface gravity $\kappa$ remains finite as $r\to R_{\rm s}^+$. However, curvature invariants such as the Ricci scalar $R$ and the Kretschmann scalar ${\cal K}$ are still divergent, making this point a truly singular point of the manifold. This singularity can be removed if we take a density function that approaches $R_{\rm s}$ in a smoother way, specifically a $\rho(r)$ of class at least $C^2$, i.e.~$\lambda>2$. In this context, where all second-derivative curvature invariants are finite at the surface, the domain of validity of the exterior metric can be extended up to $r=\Rs$, without having to restrict it to a Planck length from the star's surface, namely~$r=\Rs+\lP$, as was discussed for the constant density case in~\cite{Calmet:2019eof,Perrucci:2024qrr}. More in general, we have found that, for the exterior spacetime $g_{\mu\nu}^{\rm ext}$ to be of class $C^k$ at $r=\Rs$, $\rho(r)$ must be at least of class $C^{k+1}$, which, for the modified Tolman {\rm VII} density profile~\eqref{eq:MTVII}, corresponds to $\lambda>k+1$.

Additionally, we have studied the asymptotic behaviour of the quantum corrections far away from the star's surface, i.e. for~$r\gg\Rs$. We have found that the dependence of the exterior metric perturbations on the internal structure of the matter source, through the density parameter $\lambda$, enters at next-to-leading order in the series expansions~\eqref{eq:h00MTVIIxinf} and~\eqref{eq:h11MTVIIxinf}. Our results provide further evidence that the dependence on the interior composition of the object, also known as quantum ``hair'', enters at order $r^{-5}$, as originally observed in~\cite{Calmet:2021stu}. 
To conclude, by solving the integral in Eq.~\eqref{eq:Deltaphi}, we have computed the quantum-corrected expression for the deflection angle $\delta\phi$, which illustrates how the metric perturbations $h_{tt}^{\rm ext}$ and $h_{rr}^{\rm ext}$ enter an observable quantity. For the background spacetime of a star in stable equilibrium, however, their contribution is suppressed by a factor of $\lPs/R_{\rm s}^2$, rendering the deviation from general relativity effectively unmeasurable.


\section*{Acknowledgments}
M.S. thanks Tom Gent for pointing out a useful special-function identity and Tommaso Antonelli for a helpful comment on the deflection-angle integral. The work of M.S. is supported by a doctoral studentship of the Science and Technology Facilities Council (grant ST/X508822/1, project ref. 2753640).
\section*{Data Availability Statement}
Data sharing not applicable to this article as no datasets were generated or analysed during the current study.
\section*{Code Availability Statement}
Code/Software sharing not applicable to this article as no code/software was generated or analysed during the current study.



\bigskip{}

\end{document}